\documentclass[preprintnumbers,article,amsmath,amssymb,floatfix,10pt,prd,twocolumn,
superscriptaddress,nofootinbib]{revtex4-2}
\usepackage{bm}
\usepackage{amsfonts}
\usepackage{latexsym}
\usepackage[latin1]{inputenc}
\usepackage{graphicx}
\usepackage{amsmath}
\usepackage{palatino}
\usepackage{mathpazo}
\usepackage{textcomp}
\usepackage{float}
\usepackage{booktabs}
\usepackage{dcolumn}
\usepackage{ragged2e}
\usepackage{hyperref}
\hypersetup{colorlinks,citecolor=blue}
\hypersetup{colorlinks=true,linkcolor=red,filecolor=magenta,    urlcolor=blue}
\usepackage{amsmath}
\usepackage{xcolor}
\usepackage{orcidlink}
\usepackage{epsfig}
\usepackage{caption}
\usepackage{subcaption}
\usepackage{commath}
\def\jnl@style{\it}
\def\aaref@jnl#1{{\jnl@style#1}}

\def\aaref@jnl#1{{\jnl@style#1}}

\def\aj{\aaref@jnl{AJ}}                   
\def\apj{\aaref@jnl{ApJ}}                 
\def\apjl{\aaref@jnl{ApJ}}                
\def\apjs{\aaref@jnl{ApJS}}               
\def\apss{\aaref@jnl{Ap\&SS}}             
\def\aap{\aaref@jnl{A\&A}}                
\def\aapr{\aaref@jnl{A\&A~Rev.}}          
\def\aaps{\aaref@jnl{A\&AS}}              
\def\mnras{\aaref@jnl{Mon.~Not.~Roy.~Astron.~Soc.}}             
\def\prd{\aaref@jnl{Phys.~Rev.~D}}        
\def\prc{\aaref@jnl{Phys.~Rev.~C}}  
\def\prl{\aaref@jnl{Phys.~Rev.~Lett.}}    
\def\qjras{\aaref@jnl{QJRAS}}             
\def\skytel{\aaref@jnl{S\&T}}             
\def\ssr{\aaref@jnl{Space~Sci.~Rev.}}     
\def\zap{\aaref@jnl{ZAp}}                 
\def\nat{\aaref@jnl{Nature}}              
\def\aplett{\aaref@jnl{Astrophys.~Lett.}} 
\def\apspr{\aaref@jnl{Astrophys.~Space~Phys.~Res.}} 
\def\physrep{\aaref@jnl{Phys.~Rep.}}      
\def\physscr{\aaref@jnl{Phys.~Scr}}       
\def\commat{\aaref@jnl{Comm.~Math.~Phys.}}              
\def\science{\aaref@jnl{Science}}               
\def\cqg{\aaref@jnl{Classical Quant.~Grav.}}            
\def\jpcs{\aaref@jnl{JPCS}}                                     
\def\ijmpd{\aaref@jnl{Int.~J.~Mod.~Phys.~D}}                    
\def\grg{\aaref@jnl{Gen.~Relat.~Gravit.}}               
\def\rpp{\aaref@jnl{Rep.~Prog.~Phys.}}          
\def\npa{\aaref@jnl{Nucl.~Phys.~A}}        
\def\lrr{\aaref@jnl{Living Rev.~Rel.}}                   
\def\jcap{\aaref@jnl{J.~Cosmology Astropart.~Phys.}}    
\def\rmp{\aaref@jnl{Rev.~Mod.~Phys.}}   
\def\epjc{\aaref@jnl{Eur.~Phys.~J.~C}} 
\def\plb{\aaref@jnl{~Phy.~Lett.~B}} 
\def\mpla{\aaref@jnl{Mod.~Phy.~Lett.~A}} 
\def\arxiv{\aaref@jnl{arxiv.org}}

\allowdisplaybreaks[1]

\begin{document}
\color{black}       
\title{Cosmology with viscous generalized Chaplygin gas in $f(Q)$ gravity}

\author{Gaurav N. Gadbail\orcidlink{0000-0003-0684-9702}}
\email{gauravgadbail6@gmail.com}
\affiliation{Department of Mathematics, Birla Institute of Technology and
Science-Pilani,\\ Hyderabad Campus, Hyderabad-500078, India.}

\author{Simran Arora\orcidlink{0000-0003-0326-8945}}
\email{dawrasimran27@gmail.com}
\affiliation{Department of Mathematics, Birla Institute of Technology and
Science-Pilani,\\ Hyderabad Campus, Hyderabad-500078, India.}

\author{P.K. Sahoo\orcidlink{0000-0003-2130-8832}}
\email{pksahoo@hyderabad.bits-pilani.ac.in}
\affiliation{Department of Mathematics, Birla Institute of Technology and
Science-Pilani,\\ Hyderabad Campus, Hyderabad-500078, India.}
%

\begin{abstract}

We use the hybrid model of bulk viscosity and generalized chaplygin gas (GCG), named the viscous generalized chaplygin gas (VGCG) model, which is thought to be an alternate dark fluid of the universe. We explore the dynamics of the VGCG model in the framework of the non-metricity $f(Q)$ gravity using the functional form $f(Q)=\beta Q^n$, where $\beta$ and $n$ are arbitrary constants. For the purpose of constraining model parameters, we use recent observational datasets such as Observational Hubble data, Baryon Acoustic Oscillations, and Type $Ia$ supernovae data. According to our study, the evolution of the deceleration parameter $q$ and the equation of state (EoS) parameter $w$ show a transition from deceleration to an acceleration phase and its deviation from the $\Lambda$CDM model. \\


\end{abstract}

\date{\today}

\maketitle

\section{Introduction}

Modifications to the general theory of relativity (GR) by Einstein have been active essentially since the start of the model of gravity. The role of dark energy and the cosmological constant in explaining the advent of the accelerating universe became one of the most prominent unresolved issues in cosmology. It is now widely accepted that the universe is going through a stage of accelerated expansion, confronting the various observational evidence (for a review of these data/evidence, see References \cite{Perlmutter/1999,Riess/1998,Riess/2004,Spergel/2007,Koivisto/2006}). \\
The $\Lambda$CDM model (with an equation of state $w =-1$), often known as a cosmological constant model, is the most competitive cosmological model of dark energy. However, as evidenced by several works \cite{Hinshaw/2013,Zhao/2012}, neither theoretical nor observational considerations accurately confirm $\Lambda$CDM dark energy models, encouraging the necessity of new approaches to understanding its nature.\\
Consequently, general relativity appears to be only vaguely valid at very early and highly late times. It would be highly tempting to be able to describe fundamental cosmological phenomena like early-time inflation and late-time acceleration in a coherent manner as modified gravity effects. Additionally, modified gravity may be able to explain how dark matter had a part in the creation and development of the universe and provide a solution to the coincidence problem.\\
We can list, for instance, models following the GR construction, having null non-metricity and torsion such as $f(R)$, $f(R,T)$ theories \cite{Capo/2008,Nojiri/2007, Harko/2011,Momeni/2015}. In lieu, one can formulate theories of gravity associated to torsion $T$ (teleparallel equivalent GR) \cite{Capo/2011,Nunes/2016} and non-metricity $Q$ (symmetric teleparallel equivalent GR) \cite{Jimenez/2018}, which attempts to unify field theories \cite{Bahamonde/2021}. 
We will consider an extension of the symmetric teleparallel GR, the $f(Q)$ theory, for which gravity is attributed to the non-metricity scalar $Q$. Here, the non-metricity $Q$ represents the basic geometric variable characterizing how the length of a vector changes when transported. Detailed investigations of this theory have been performed in many directions \citep{Harko/2018,Mandal/2020,Frusciante/2021,Solanki/2021,Khyllep/2021,Ayuso/2021,GG/2022,Capo/2022,Wang/2022}.\\
The chaplygin gas (CG) \cite{Chaplygin/1904} had an origin that was not cosmological, but its negative pressure has recently attracted new interest in cosmology. The model explains the cosmic expansion history of a universe filled with an exotic fluid when combined with the FLRW background in GR. The chaplygin gas (CG) model is one of the models with a non-constant EoS parameter, which includes some exciting characteristics \cite{Setare/2007,Bamba/2012}. In fact, the CG models have been recently investigated in the context of modified gravity theories \cite{Saha/2019,Sahlu/2019,Jamil/2008}. But the CG model has some issues with structure formulation and cosmological power spectrum \cite{Bean/2003,Sandvik/2004}.\\
Lately,  Bento et al. \cite{Bento/2002} presented generalized chaplygin gas (GCG) model to address or mitigate the CG-related issues with dark energy. It is known that the GCG model has undergone extensive research to explain the accelerating universe and has been supported by observations \cite{Bento/2003,Bertolami/2004,Barreiro/2008,Bento/2003a,Yilmaz/2021}. 
The GCG's remarkable feature is its ability to characterize dark matter and dark energy in the context of a unified fluid through an exotic equation of state \cite{Makler/2003,Amendola/2003}. The GCG is expressed by the equation of state $p=-\frac{A}{\rho^{\alpha}}$, where $A>0$ and $0<\alpha\leq 1$. It is noted that $\alpha=1$ corresponds to the CG scenario \cite{Kamenshchik/2001}, and $\alpha=0$ with $A=1$ reduces to the $\Lambda$CDM.\\
Another case of DE that might play an essential role in the evolution of the universe has to do with the dissipative phenomena in the form of shear and bulk viscosity. Thereafter, the bulk viscosity has been investigated concerning inflation and as a possible cause of the accelerated expansion of the universe in the Eckart formalism \cite{Eckart/1940,Barrow/1987,Normann/2017,Brevik/2018}. The bulk viscosity redefines the effective pressure as $p_{ eff}=p-3 H \zeta$,  where $H$ is the Hubble parameter and $\zeta$ is the anticipated bulk viscous coefficient based on the background geometry.  Typically, the widely investigated power-law form of the density-dependent bulk viscous coefficient has the form $\zeta(\rho)=\zeta_0\rho^{\lambda}$ with  $\zeta_0>0$, ensuring a positive entropy \cite{Zimdahl/2000,Cardenas/2015,Normann/2016}. \\
Here, we use the hybrid model of bulk viscosity and GCG, named the viscous generalized chaplygin gas (VGCG) model initially proposed in \cite{Zhai/2006}, which is widely studied to describe the observed accelerated expansion of the universe. The VGCG model explains the universe's acceleration through a unique equation of state that produces behavior similar to that of dark matter at early times and dark energy at later times. However, the fundamental issue with such a unified model is that it now causes nonphysical oscillations or exponential blowups in the matter power spectrum. Additionally, it was demonstrated that the linear approximation fails early on, indicating the need for a more cautious approach that considers the nonlinear effects \cite{ Sandvik/2004}. That is, rather than an energy density with a GCG equation of state. We assume that a modified kind of gravity causes the background evolution. Inspired by the above discussion and investigations in modified theories of gravity, the idea is to study the generalized Chaplygin gas interacting with $f(Q)$ gravity in the presence of bulk viscosity fluids in this framework. Several authors have investigated the VGCG model in the framework of GR and modified gravity to avoid causality problems that arise when dissipative fluids are considered solely \cite{Xu/2012,Amani/2013,Lia/2015,Almada/2021,Baffou/2017}. There are other works done in VGCG that are not confronted with observational data \cite{Debnath/2015,Rudra/2013}. The present work analyzes the hybrid model: VGCG in the framework of $f(Q)$ gravity theory. Firstly, we produce the feasible solutions that rescue to $\Lambda$CDM scenario to the approximate cosmological limits using the VGCG equation of state. Furthermore, we perform Bayesian joint analysis using the following samples: observational Hubble parameter (OHD),  Type Ia supernovae (SNe Ia), and baryon acoustic oscillations (BAO).\\
The manuscript is organized as follows: Section \ref{section 2} provides the basic formalism of the $f(Q)$ gravity. Section \ref{section 3} presents the viscous generalized chaplygin gas model in the $f(Q)$ gravity framework and the expressions for the Hubble parameter. Section \ref{section 4} summarizes of the cosmological data and presents the methodology to establish bounds on the model parameters. In section \ref{section 5}, we study the evolution of cosmological parameters. Lastly, we conclude our results in section\ref{section 7}.

\section{$f(Q)$ theory of gravity}
\label{section 2}
The action of the $f(Q)$ gravity is considered as \cite{Jimenez/2018}
\begin{equation}
\label{1}
S=\int \left[\frac{1}{2}f(Q)+\mathcal{L}_m\right]\sqrt{-g}d^4x,
\end{equation}
where $\mathcal{L}_m$ is the matter Lagrangian density, $f(Q)$ is an arbitrary function of non-metricity scalar $Q$ and $g$ is determinant of the metric tensor $g_{\alpha\beta}$. Here, we assume $8\pi G =1$. The nonmetricity tensor in $f(Q)$ gravity is defined as $Q_{\sigma\alpha\beta}=\nabla_{\sigma}\,g_{\alpha\beta}$ and the corresponding  traces are 
\begin{equation}
\label{2}
Q_{\sigma}=Q_{\sigma\,\,\,\,\alpha}^{\,\,\,\,\alpha}\, ,\,\,\,\,\,\,\,\,\tilde{Q}_{\sigma}=Q^{\alpha}_{\,\,\,\,\sigma\alpha}\,.
\end{equation}
Additionally, the non-metricity conjugate or the superpotential tensor $P_{\,\,\mu\nu}^{\lambda}$ is given by
\begin{equation}
\label{3}
4P_{\,\,\alpha\beta}^{\sigma}=-Q^{\sigma}_{\,\,\,\,\alpha\beta}+2Q^{\,\,\,\,\,\,\sigma}_{(\alpha\,\,\,\,\beta)}-Q^{\sigma}g_{\alpha\beta}-\tilde{Q}^{\sigma}g_{\alpha\beta}-\delta^{\sigma}_{(\alpha}\, Q\,_{\beta)},
\end{equation} 
acquiring the nonmetricity scalar as 
\begin{equation}
\label{4}
Q=-Q_{\sigma\alpha\beta}P^{\sigma\alpha\beta}.
\end{equation}
The energy-momentum tensor for matter read as
\begin{equation}
\label{5}
T_{\alpha\beta}\equiv-\frac{2}{\sqrt{-g}}\frac{\delta(\sqrt{-g})\mathcal{L}_m} {\delta g^{\alpha\beta}}.
\end{equation}
The gravitational field equation of $f(Q)$ gravity obtained by varying action \eqref{1} with respect to the metric is written as
\begin{multline}
\label{6}
\frac{2}{\sqrt{-g}}\nabla_{\sigma}\left(f_{Q}\sqrt{-g}\,P^{\sigma}_{\,\,\alpha\beta}\right)+\frac{1}{2}f\,g_{\alpha\beta}+\\
f_{Q}\left(P_{\alpha\sigma\lambda}Q_{\beta}^{\,\,\,\sigma\lambda}-2Q_{\sigma\lambda\alpha}P^{\sigma\lambda}_{\,\,\,\,\,\,\beta}\right)=- T_{\alpha\beta},
\end{multline}
where $f_Q=\frac{d f}{d Q}$. Likewise, varying equation \eqref{1} with respect to the connection gives:
\begin{equation}
\label{7}
\nabla_{\alpha}\nabla_{\beta} \left(f_{Q}\sqrt{-g}\,P_{\,\,\,\,\,\,\,\sigma}^{\alpha\beta}\right)=0.
\end{equation}

The spatially-flat Friedmann-Lemaitre-Robertson-Walker (FRW) metric, describing the homogeneous and isotropic universe, is chosen as  
\begin{equation}
\label{8}
ds^2=-dt^2+a^2(t)(dx^2+dy^2+dz^2).
\end{equation}
Here, $a(t)$ is a cosmic scale factor and $H=\frac{\dot{a}}{a}$. In this context, we have the nonmetricity scalar $Q=6H^2$ and the energy-momentum tensor of a perfect fluid $T_{\alpha\beta}=(p+\rho)u_{\alpha}u_{\beta}+pg_{\alpha\beta}$, where $p$ and $\rho$ are the pressure and the energy density, respectively.\\ Using Eq. \eqref{8} in Eq. \eqref{6}, we find the Friedmann-like equations for $f(Q)$ as follows \cite{Jimenez/2018}: 
\begin{equation}
\label{9}
6f_Q H^2-\frac{1}{2}f=\rho,
\end{equation} 
\begin{equation}
\label{10}
(12f_{QQ}H^2+f_Q)\dot{H}=-\frac{1}{2}(p+\rho).
\end{equation}

\section{Viscous GCG model}
\label{section 3}
In this framework, we consider $f(Q)=\beta Q^n$ as an  algebraic polynomial function in $Q$ of degree $n$, where $\beta$ is a constant. It is worth noting that $\beta=n=1$ corresponds to a case of the successful theory of general relativity (GR), but not $\Lambda$CDM since the cosmological constant is absent; thus, this model alleviates the cosmological constant problem. Moreover, the model suggests small deviations from the $\Lambda$CDM model as the redshift increases \cite{Capozziello/2022}. This model is capable of describing late-time universe acceleration and is also compatible with BBN limitations \cite{Fotios/2023}.
Using the above functional form, we rewrite Eq. \eqref{9} as
\begin{equation}
\label{11}
H=\left[\frac{2\rho}{\beta (2n-1)6^n}\right]^{\frac{1}{2n}}.
\end{equation}
We construct $f(Q)$ gravity with three matters: baryonic $(b)$, radiation $(r)$ and generalized chaplygin gas $(c)$ in the presence of viscous fluid with equation of state $w_b=0$, $w_r=\frac{1}{3}$ and $\tilde{p}=-\frac{A}{\rho^{\alpha}_c}-\prod$, where $\prod=3\zeta H$, $\zeta$ is the bulk viscous coefficient, respectively. Then continuity equations for each fluid is expressed as 
\begin{eqnarray}
\label{12}
\dot{\rho}_b+3\rho_b H & = & 0,\\ 
\label{13}
\dot{\rho}_r+4\rho_r H &=&0,\\
\label{14}
\dot{\rho}_c+3\left(\rho_c+\tilde{p} \right) H&=&0.
\end{eqnarray}  
Here,  the dot represents the derivative with respect to time. The integration of Eq. \eqref{12} and \eqref{13} yields $\rho_b=\rho_{0b}\,a^{-3}$ and $\rho_r=\rho_{0r}\,a^{-4}$. 
To integrate the one for the VGCG we assume $\zeta (\rho)=\zeta_0\,\rho_c^{\lambda}$ \cite{Cardenas/2015,Normann/2016}, where $\lambda=\left(1-\frac{1}{2n}\right)\geq0$ and $\zeta_0$ is a viscous constant. For any power of $\lambda$, the above ansatz has been often utilized in the literature, both for the early and late universe. The two most common values for $\lambda$ are $\lambda=1/2$, which results in $\zeta \propto \sqrt{\rho}$, and $\lambda=1$, which results in $\zeta \propto \rho$. In this work, we use the power-law functional form of $f(Q)$ model with power $n$ and then introduce a functional form for the bulk viscous coefficient (again, a power-law). In order to make a reasonable assumption, we relate these two models by using the power in terms of $n$. The value of $\lambda$ is $\lambda\approx \frac{1}{2}$  based on our observational value of $n$,  which has already been investigated in some literature \cite{Li/2010,Brevik/2005,Brevik/2006}. Hence, we have $H^2\approx\left[\frac{2\rho_c}{\beta (2n-1)6^n}\right]^{\frac{1}{2n}}$ \footnote{A good approximation for the late stage of the Universe} and
\begin{equation}
\label{15}
\rho_c=\rho_{0c}\left[k+(1-k)a^{-3(1-\mathcal{X})(\alpha+1)}\right]^{\frac{1}{\alpha+1}}.
\end{equation}
 We define $\mathcal{X}=3\zeta_0\left(\frac{2}{(2n-1)\beta 6^n}\right)^{\frac{1}{2n}}$ and $k=\frac{A}{\rho^{\alpha+1}_{0c}\mathcal{X}}$. Rewriting Eq.\eqref{11} as follows: 
\begin{equation}
\label{16}
H=\left[\frac{2(\rho_b+\rho_r+\rho_c)}{\beta (2n-1)6^n}\right]^{\frac{1}{2n}}.
\end{equation}
Hence, using $\rho_b$, $\rho_r$, and $\rho_c$, one gets the Hubble parameter $H$ in terms of cosmic scale factor $(a)$ as

\begin{multline}
\label{17}
H(a)=H_0^{\frac{1}{n}} \left(\frac{6^{1-n}}{(2n-1)\beta}\right)^{\frac{1}{2n}}\left[\Omega_{0b}\, a^{-3}+\Omega_{0r}\, a^{-4} + \right. \\
\left. \Omega_{0c}\left(k+(1-k)a^{-3(1-\mathcal{X})(\alpha+1)}\right)^{\frac{1}{\alpha+1}} \right]^{\frac{1}{2n}}.
\end{multline}

Furthermore, the Hubble parameter in terms of redshift $z$ is obtained by using the relation $a=\frac{1}{z+1}$

\begin{multline}
\label{18}
H(z)=H_0^{\frac{1}{n}} \left(\frac{6^{1-n}}{(2n-1)\beta}\right)^{\frac{1}{2n}}\left[\Omega_{0b}\,(1+z)^{3}+\Omega_{0r}\,(1+z)^{4} \right. \\ \left. +\Omega_{0c}\left((1-k)(1+z)^{3(1-\mathcal{X})(\alpha+1)}
+k \right)^{\frac{1}{\alpha+1}} \right]^{\frac{1}{2n}}.
\end{multline}

Imposing the initial condition $H(0)=H_0$ in the above equation, we have the following constraint:
\begin{equation}
\label{19}
\Omega_{0c}=H_0^{2n-2}6^{n-1}(2n-1)\beta-\Omega_{0b}-\Omega_{0r}
\end{equation}

\section{Methodology}
\label{section 4}
The following are the most recent and relevant observational results that we used in our analysis:

\begin{itemize}
\item Observational Hubble data (OHD): We consider the $H(z)$ data points measured by calculating the differential ages of galaxies as a function of redshift $z$  \cite{Yu/2018,Moresco/2015,Sharov/2018}.
\item  Baryon acoustic oscillation (BAO): We also consider the BAO measurements by SDSS-MGS, Wiggle Z, and 6dFGS collaborations \cite{Blake/2011,Percival/2010,Giostri/2012}.
\item Type-Ia Supernova measurement (SNe Ia): We consider the
Pantheon sample of 1048 SNe Ia measurements of luminosity distance from the Pan-STARSS1 (PS1) Medium Deep Survey, the Low-z, SDSS, SNLS, and HST surveys \cite{Scolnic/2018,Chang/2019}.
\end{itemize}

To establish the joint constraints on parameter space $(\alpha, \beta, n, \zeta_{0}, k, \Omega_{b})$ from the aforementioned cosmic probe, we employ the total likelihood function. The best fits of the parameters are maximized by using the probability function $\mathcal{L} \propto e^{-\chi^2/2}$ \cite{Lazkoz/2019,Anag/2021}
\begin{equation*}
\mathcal{L}_{tot}=  \mathcal{L}_{OHD} \times \mathcal{L}_{BAO} \times \mathcal{L}_{SNe\,Ia}.
\end{equation*}
Henceforth, the corresponding $\chi^{2}$ reads 
\begin{equation*}
    \chi^{2}_{tot}=   \chi^{2}_{OHD} + \chi^{2}_{BAO}+ \chi^{2}_{SNe\,Ia}.
\end{equation*}
Finally, we adopt the Markov Chain Monte Carlo (MCMC) sample from the python package $emcee$ for likelihood minimization, frequently used in astrophysics and cosmology to explore the parameter space.\\
Figure \ref{combine69} and \ref{combine73} show the results of $1-\sigma$ and $2-\sigma$ likelihood contours with the best-fit values of model parameters. We consider $H_{0}= 69$ $Km/s/Mpc$ \cite{Planck/2018} and $H_{0}=73.2$ $Km/s/Mpc$ \cite{Valentino/2021}, $\Omega_{r}=0.00005$ in our analysis. Furthermore, the error bars for $H(z)$ and $\mu(z)$ are shown in figures \ref{errorHub} and \ref{errorMu} using $H_{0}= 69$ $Km/s/Mpc$. \\
It is observed that the model with $H_{0}=69$ and $73.2$ yields $\chi^2$ values as $1089.11$ and $1112.91$, respectively. The $\chi^2$ for the $\Lambda$CDM is obtained as $1059.9$.

\begin{widetext}

\begin{figure}[H]
\centering
\includegraphics[scale=0.55]{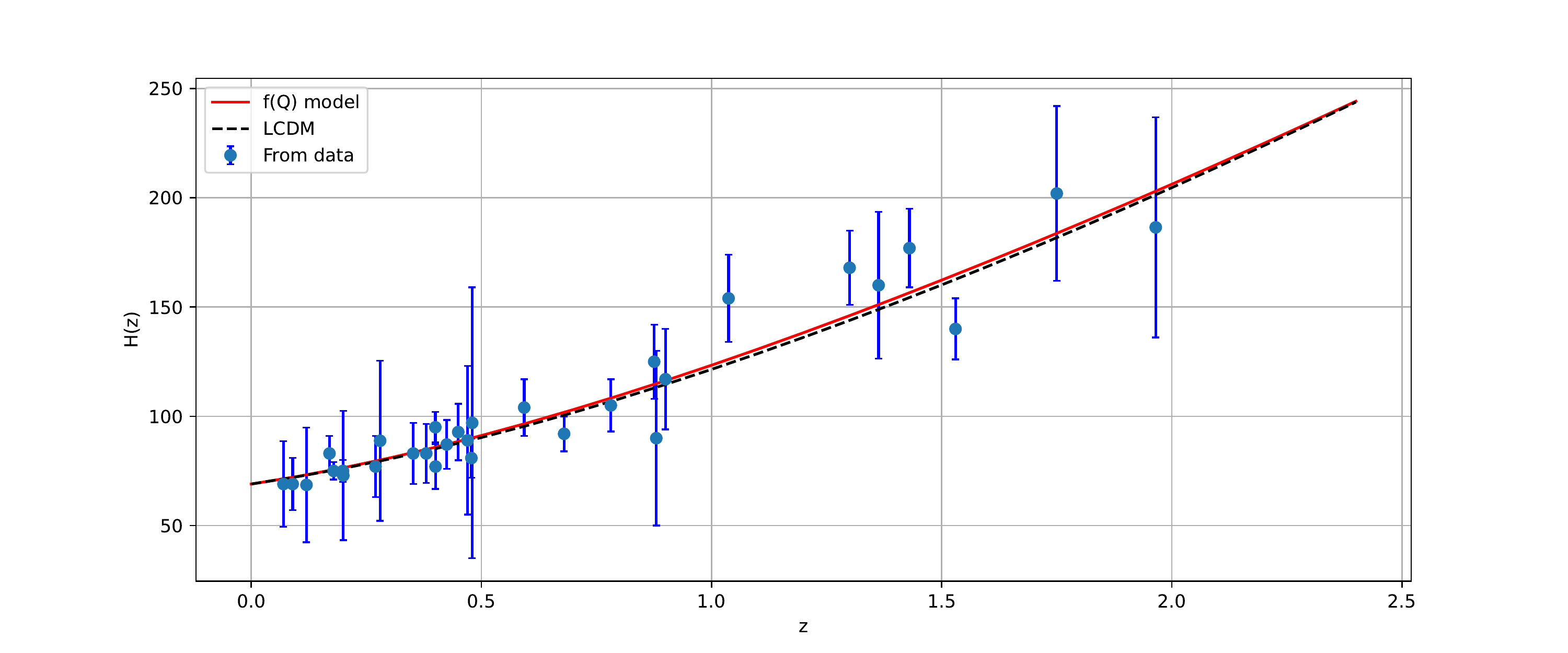}
\caption{The Hubble parameter evolution with respect to redshift $z$. The black dashed line corresponds to the $\Lambda$CDM model, the red line is the curve obtained for our model, and the blue dots represent error bars. }
\label{errorHub}
\end{figure}

\begin{figure}[H]
\centering
\includegraphics[scale=0.45]{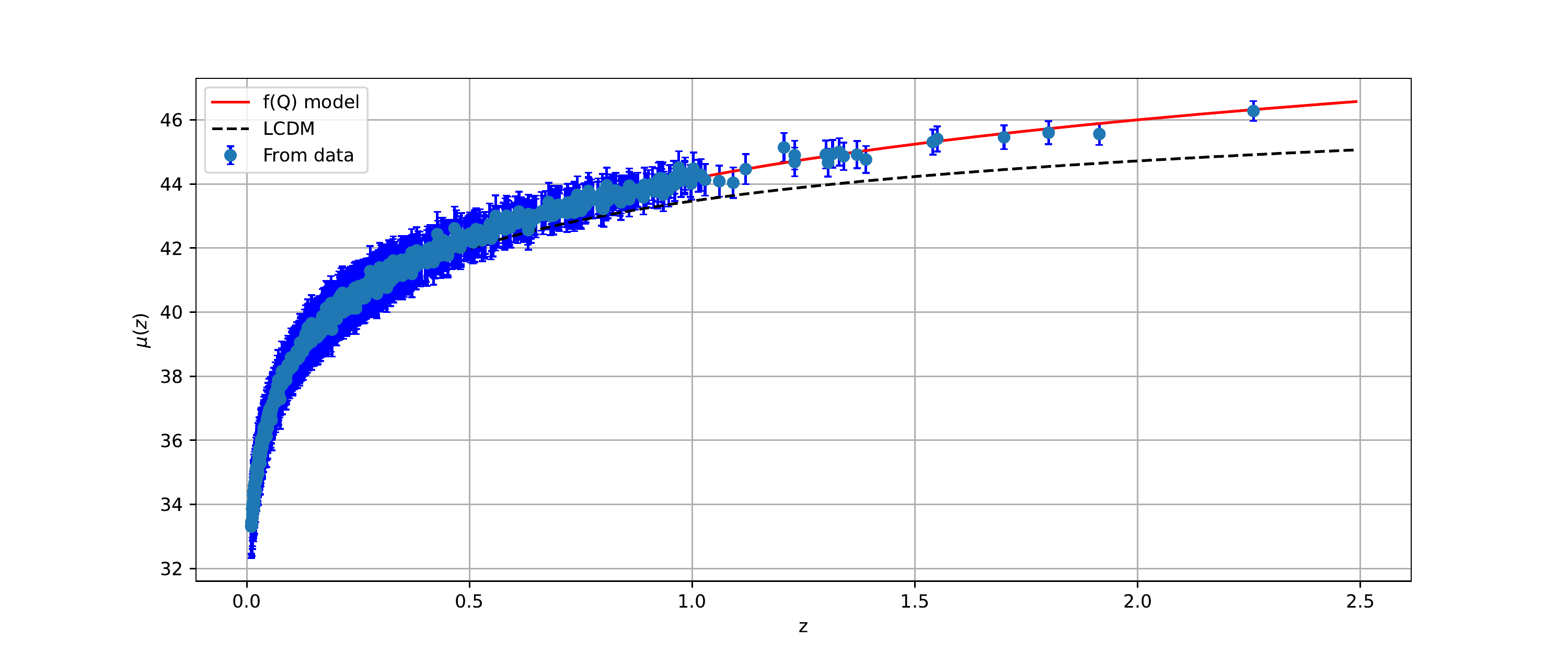}
\caption{The plot of $\mu (z)$ with respect to redshift z. The black dashed line corresponds to the $\Lambda$CDM model, the red line is the curve obtained for our model, and the blue dots represent error bars.}
\label{errorMu}
\end{figure}

\begin{figure}[H]
\centering
\includegraphics[scale=0.47]{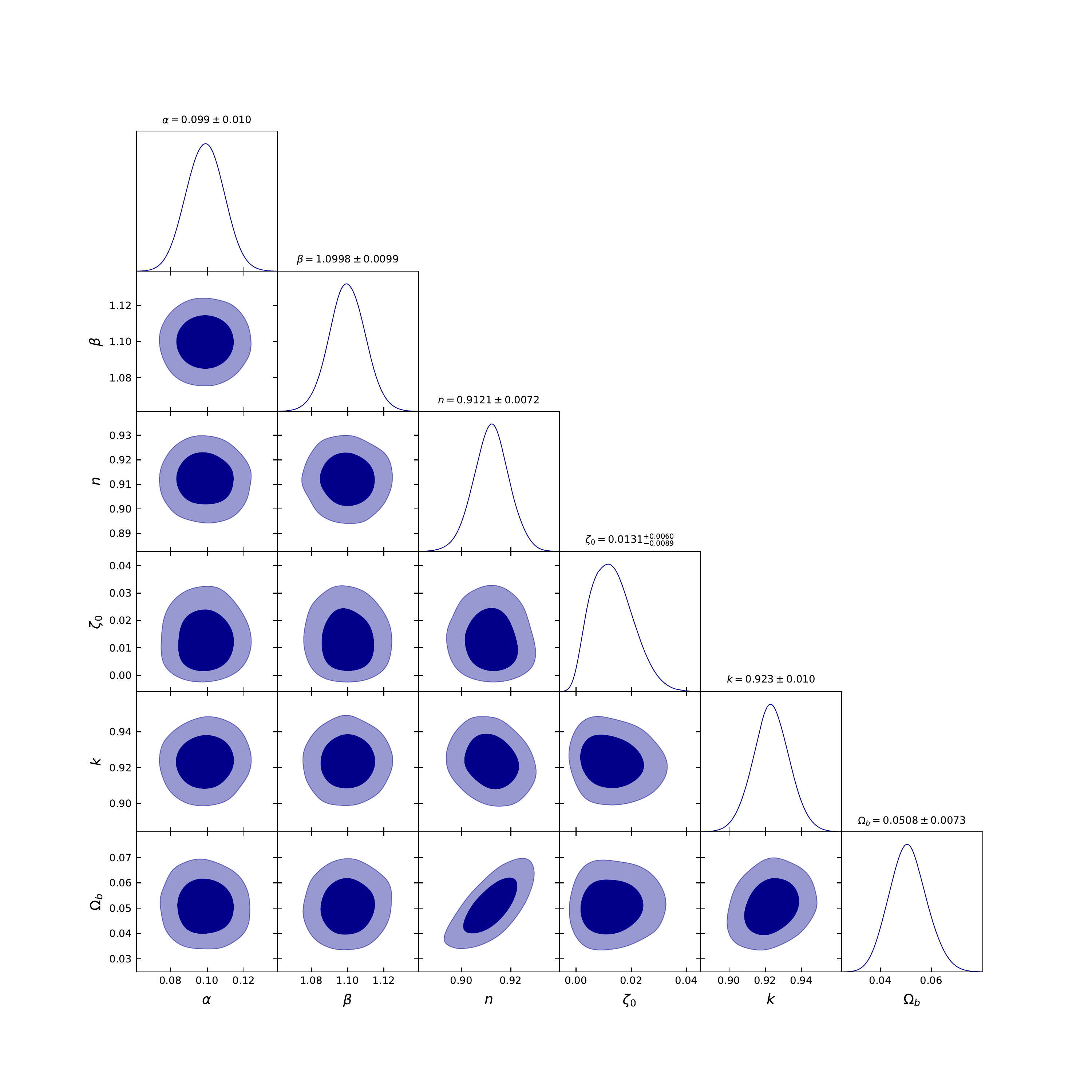}
\caption{The $1-\sigma$ and $2-\sigma$ confidence regions of $(\alpha, \beta, n, \zeta_{0}, k, \Omega_{b})$ for $H_{0}= 69$ $Km/s/Mpc$ using $OHD+BAO+SNe\,Ia$. }
\label{combine69}
\end{figure}

\begin{figure}[H]
\centering
\includegraphics[scale=0.5]{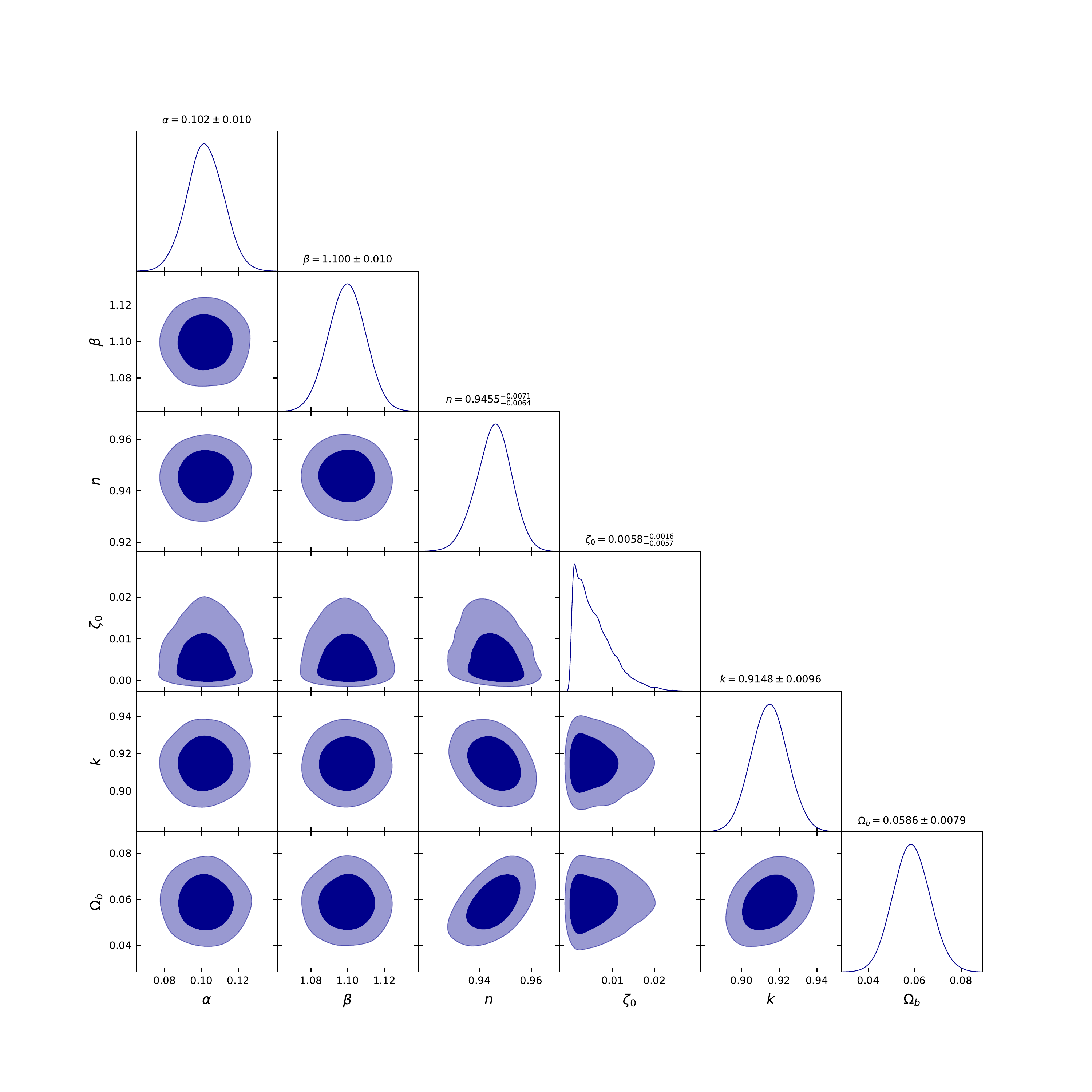}
\caption{The $1-\sigma$ and $2-\sigma$ confidence regions of $(\alpha, \beta, n, \zeta_{0}, k, \Omega_{b})$ for $H_{0}= 73.2$ $Km/s/Mpc$ using $OHD+BAO+SNe\,Ia$.}
\label{combine73}
\end{figure}

\end{widetext}

\section{Evolution of cosmological parameters}
\label{section 5}

\begin{figure}[H]
\centering
\includegraphics[scale=0.6]{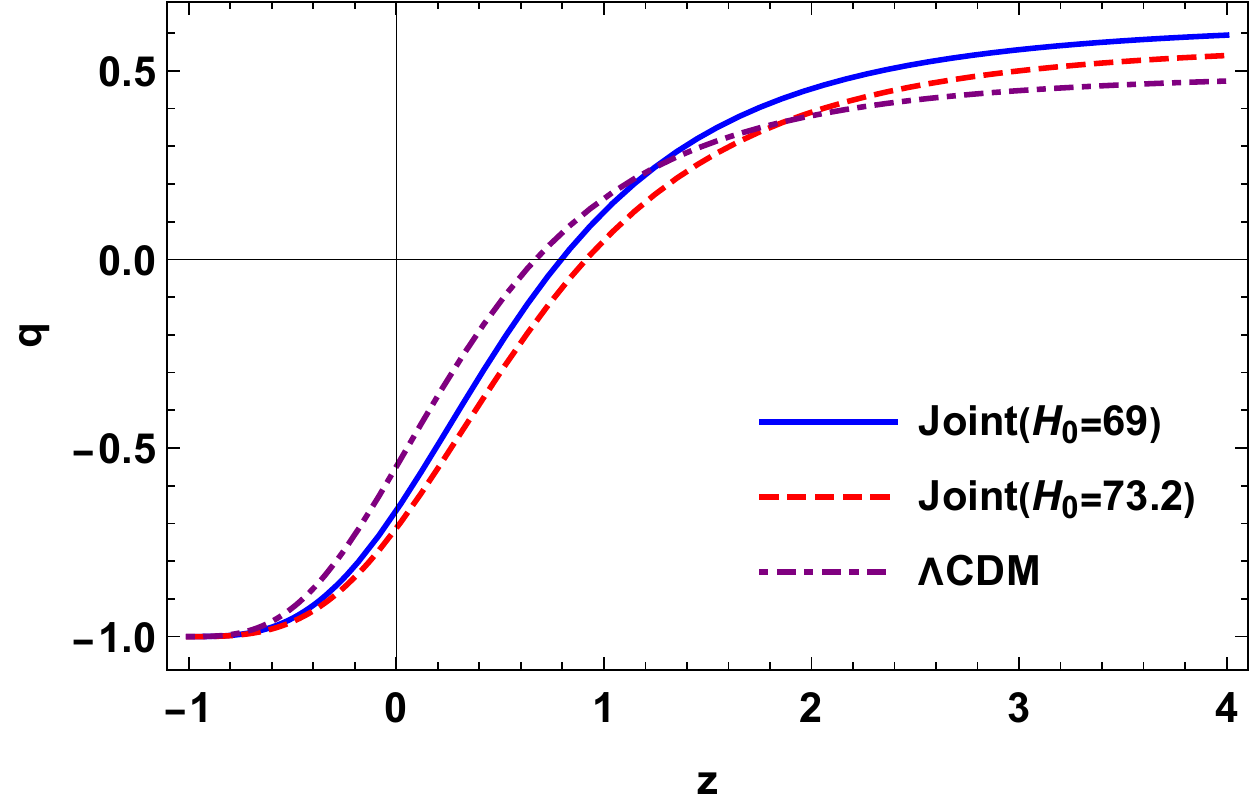}
\caption{Evolution of the deceleration parameter $q$ versus redshift $z$.}
\label{q}
\end{figure}

In this section, we would like to study the evolution of various cosmological parameters and compare our estimated values with observational data.\\
Theoretically, the universe should decelerate in the absence of dark energy, since gravity holds matter together. In order to adequately describe the whole evolutionary history of the universe, a cosmological model requires both decelerated and accelerated expansion phases. The deceleration parameter plays an essential role in this context, and it is defined as follows:
\begin{equation}
q=-\frac{a\ddot{a}}{\dot{a}^2}=-1-\frac{\dot{H}}{H^2}
\end{equation} 
The parameter $q$ can be positive as well as negative. So, $q>0$ indicates that the universe is under decelerating phase, and there is a domination of matter over dark energy. However, $q<0$ implies the accelerating phase and domination of dark energy.\\
To show that the model can account for the acceleration in the universe, we plot the deceleration parameter $q$ in figure \ref{q}.
The deceleration parameter starts at high positive values and acceleration emerges with $q$ passing zero at redshift $z_t=0.79^{+0.02}_{-0.02}$ and $z_t=0.9^{+0.015}_{-0.025}$. Note that the present value of $q$ is $q_0=-0.66^{+0.014}_{-0.100}$ and $q_0=-0.71^{+0.15}_{-0.20}$ \cite{Santos/2016,Mamon/2018,Mehrabi/2021} corresponding to the constrained values of model parameters by $OHD+SNe\,Ia+BAO$ dataset for $H_0=69$ $Km/s/Mpc$ and $H_0=73.2$ $Km/s/Mpc$, respectively. Additionally, the evolution of $q$ for VCGC model is comparable with $\Lambda$CDM at intermediate redshifts ($z \sim 0.5-1.3$) with the consistent values at $\sigma$ confidence levels.

\begin{figure}[]
\centering
\includegraphics[scale=0.6]{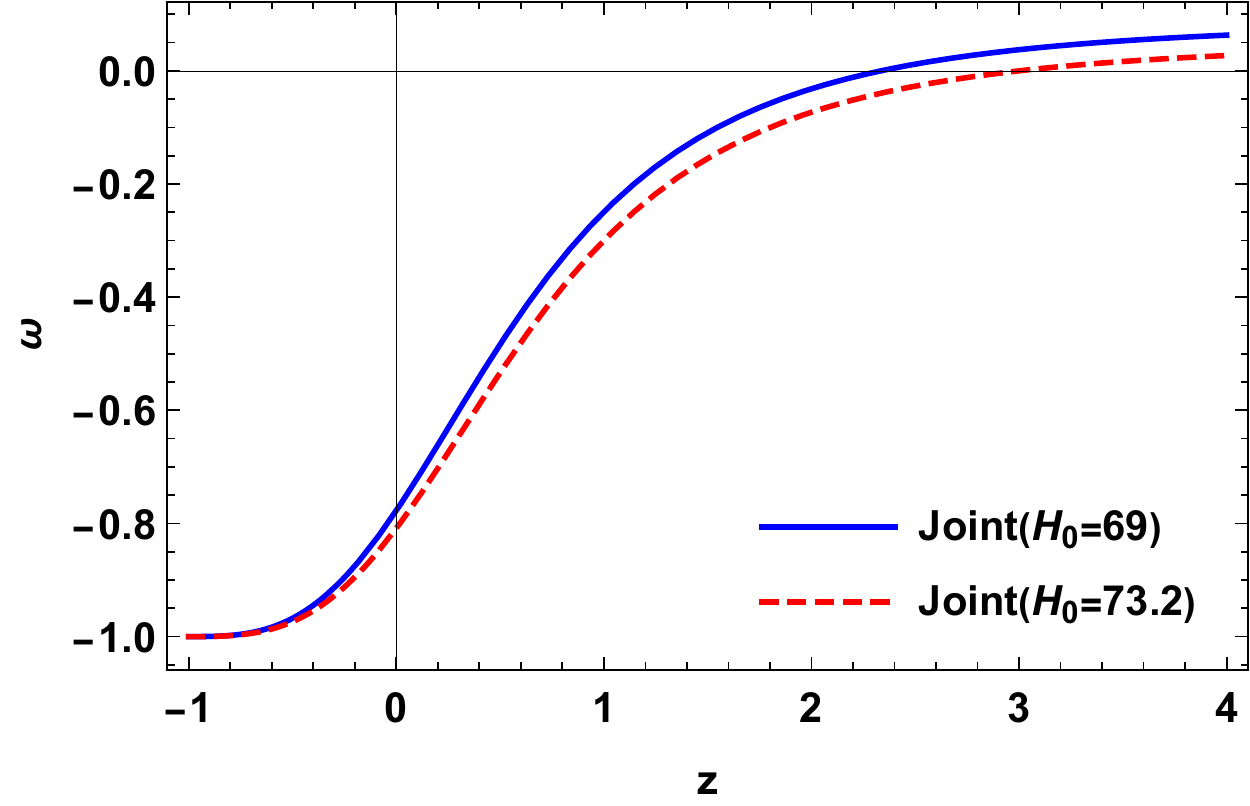}
\caption{Trajectory of the effective EoS $w_{eff}$ versus redshift $z$.}
\label{w}
\end{figure}

\begin{figure}[]
\centering
\includegraphics[scale=0.6]{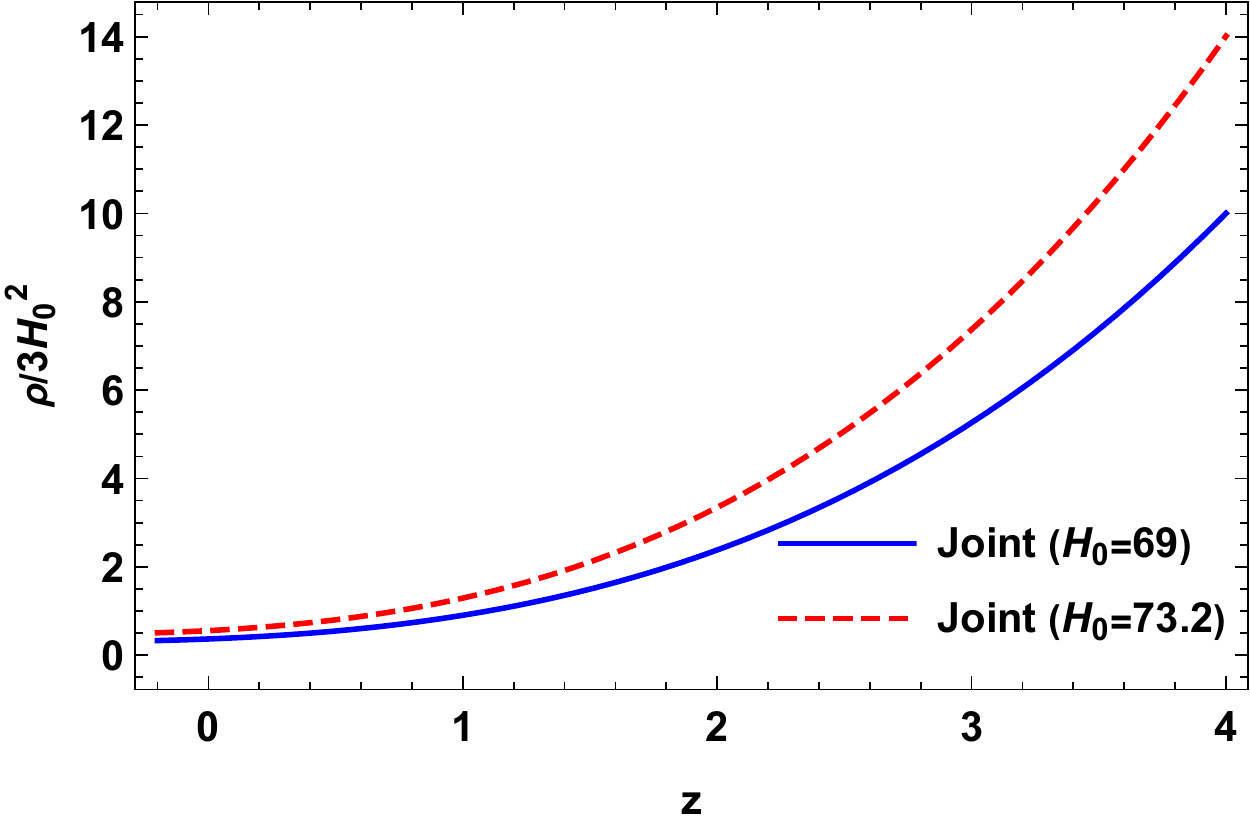}
\caption{Trajectory of the density $\frac{\rho}{3H_0^2}$ versus redshift $z$.}
\label{rho}
\end{figure}

Furthermore, the effective equation of state parameter $w = \frac{p}{\rho}$ is presented in figure \ref{w}. Recall that the $w<-\frac{1}{3}$, which includes the quintessence $(-1<w\leq 0)$ and phantom phantom $(w<-1)$ regimes, is the prerequisite for an accelerating universe. Moreover, the EoS parameter characterize vacuum energy or the cosmological constant as $w=-1$. \\
It is noted that the cosmic viscous fluid has characteristics of a quintessence dark energy. The present value of the EoS parameter from the combined dataset is obtained as $w_{0}=-0.78^{+0.07}_{-0.006}$  and $w_{0}=-0.81^{+0.12}_{-0.09}$ \cite{Almada/2019,Arora/2021} for $H_0=69$ $Km/s/Mpc$ and $H_0=73.2$ $Km/s/Mpc$, respectively.
The parameter $w$ rapidly decreases and eventually converges to $-1$ at a late time. \\
The density parameter behaves positively for all constrained values of model parameters as expected, as shown in figure \ref{rho}.

\section{Conclusion}
\label{section 7}
We focused on a gravitational action that involves a general function of non-metricity scalar, which gives rise to the class of $f(Q)$ theories of gravity. So, we propose a hybrid VGCG model in the $f(Q)$ gravity approach, using the good observational fit of a VGCG model to the background evolution of the universe as a motivation. The strength of this model is its capability or  reproducing the dynamics of the universe without the necessity for a DE component of unknown nature. Fot this purpose, we constrained model parameters using the MCMC method with the joint analysis of $OHD$, $SNeIa$, and $BAO$ data, as illustrated in the fourth section of this study.\\
Based on our joint analysis, the other section was dedicated to the evolutionary trajectory of the deceleration, EoS, and density parameters. It is worth noting that the deceleration parameter experiences the transition from a deceleration phase to the acceleration phase with transition redshift $z_t=0.79^{+0.02}_{-0.02}$ and $z_t=0.9^{+0.015}_{-0.025}$ and we obtained the present value of $q$ as $q_0=-0.66^{+0.014}_{-0.100}$ and $q_0=-0.71^{+0.15}_{-0.20}$ by considering $H_0=69$ $Km/s/Mpc$ and $H_0=73.2$ $Km/s/Mpc$, respectively. One can observe that the deceleration parameter experiences an early transition of the universe when $H_{0}=73.2$ $Km/s/Mpc$ and the best fit value of $\zeta_{0}$ is decreasing. On the other hand, we found a quintessence behavior for the hybrid model in $f(Q)$ at present with $w_{0}=-0.78^{+0.07}_{-0.006}$  and $w_{0}=-0.81^{+0.12}_{-0.09}$. It is critical to note that even the more exact observational data specify the EoS parameter $w$ to be somewhat different from $-1$.\\
Finally, our study concludes that our VGCG model can efficiently describe the late-time cosmic acceleration in the $f(Q)$ gravity framework. 

\section*{Data Availability Statement}
There are no new data associated with this article.

\section*{Acknowledgments}
GNG acknowledges University Grants Commission (UGC), New Delhi, India for awarding Junior Research Fellowship (UGC-Ref. No.: 201610122060). SA acknowledges BITS-Pilani, Hyderabad Campus for Institute fellowship. PKS acknowledges IUCAA, Pune, India for providing support through the visiting Associateship program and Science and Engineering Research Board, Department of Science and Technology, Government of India for financial support to carry out the Research project No.: CRG/2022/001847. We are very much grateful to the honorable referees and to the editor for the
illuminating suggestions that have significantly improved our work in terms
of research quality, and presentation.

\section*{Appendix: Data fitting method}
\subsection{OHD}
We adopt the widely used compilation through the differential age (DA) method, which allows us to estimate the expansion rate of the universe at redshift $z$. Thus, $H(z)$ can be predicted using 
\begin{equation*}
    H(z)= -\frac{1}{1+z} \frac{dz}{dt}.
\end{equation*}
The chi-square ($\chi^{2}$) for OHD is calculated as follows:
\begin{equation*}
    \chi^{2}_{OHD} = \sum_{i=1}^{31}  \frac{\left[H(\theta, z_{i})- H_{obs}(z_{i})\right]^2}{\sigma(z_{i})^2},
\end{equation*}
where  $H(z_{i})$ is the theoretical value for a given model at redshifts $z_{i}$, and $\theta$ is the parameter space, $H_{obs}(z_{i})$ is the observational value, $\sigma(z_{i})$ represents the observational error.

\subsection{BAO}
In the case of BAO data, we use a compilation from SDSS, 6dFGS, and Wiggle Z surveys at various redshifts. This work includes BAO data and the following cosmology as
\begin{eqnarray*}
d_{A}(z) & = & c \int_{0}^{z} \frac{d\tilde{z}}{H(\tilde{z})},\\
D_{v}(z) & = & \left[ \frac{d_{A}^2 (z) c z }{H(z)} \right]^{1/3},
\end{eqnarray*}
where $d_{A}(z)$ is the comoving angular diameter distance, and $D_{v}$ is the dilation scale. The $\chi^{2}$ for BAO is taken as
\begin{equation*}
    \chi^{2}_{BAO} = Y^{T} C_{BAO}^{-1} Y.
\end{equation*}
Here, Y depends on the considered survey and $C_{BAO}$ is the
covariance matrix \cite{Giostri/2012}.

\subsection{SNe Ia}

To find the best values using SNe Ia, we start with the observed distance modulus $\mu_{obs}$ derived from the SNe Ia detections and measure its difference from the theoretical value $\mu_{th}$. The present study accounts for the Pantheon sample, a recent SNe Ia dataset with 1048 points of distance moduli $\mu_{obs}$ at various redshifts in the range $0.01<z<2.26$. \\
The distance modulus of each supernova can be estimated using the equations:
\begin{eqnarray*}
  \mu_{th}(z) &=& 5 log_{10} \frac{d_{l}(z)}{Mpc}+25, \\
  d_{l}(z) &=& c(1+z) \int_{0}^{z} \frac{dy}{H(y,\theta)}.
\end{eqnarray*}
The distance modulus can be derived from the relation 
\begin{equation*}
    \mu= m_{B}-M_{B}+\alpha x_{1} - \beta\, c + \Delta_{M} + \Delta_{B},
\end{equation*}
where $m_{B}$ denotes the observed peak magnitude at the B-band maximum, and $M_{B}$ indicates the absolute magnitude. The parameters $c$, $\alpha$, $\beta$, and $x_{1}$ refer to the color at the brightness point, the luminosity stretch-color relationship, and light color shape, respectively. Further, $\Delta_{M}$ and $\Delta_{B}$ are distance corrections based on the mass of the host galaxy and simulation-based predicted biases.\\
The nuisance parameters in the above formula were retrieved using a new approach termed BEAMS with Bias Corrections (BBC) \cite{Kessler/2017}. Hence, the observed distance modulus reduces to the difference between the apparent magnitude $m_{B}$ and the absolute magnitude $M_{B}$, i.e., $\mu = m_{B}-M_{B}$.\\
The corresponding $\chi^{2}$ is defined as
\begin{equation*}
    \chi^{2}_{SN} = min \sum_{i,j=1} ^{1048} \Delta \mu_{i} \left( C_{SN}^{-1}\right)_{ij} \Delta_{j},
\end{equation*}
where $\Delta= \mu_{th}-\mu_{obs}$ and $C_{SN}$ is the covariance matrix.

\end{document}